\documentclass[preprint2]{aastex}

\slugcomment{Accepted for publication in the AJ; submitted 2023 September 8, accepted October 23.}

\shorttitle{Variability in PPNe: X. Long-P}

\shortauthors{Hrivnak et al.}

\begin{document}

\title{Variability in Protoplanetary Nebulae: X. Multi-year Periods as an Indicator of Potential Binaries} 

\author{Bruce J. Hrivnak\altaffilmark{1,2},  Wenxian Lu\altaffilmark{1}, Gary Henson\altaffilmark{2,3}, and Todd C. Hillwig\altaffilmark{1,2}}

\altaffiltext{1}{Department of Physics and Astronomy, Valparaiso University, 
Valparaiso, IN 46383, USA; bruce.hrivnak@valpo.edu, todd.hillwig@valpo.edu, wen.lu@valpo.edu (retired)}
\altaffiltext{2}{Southeastern Association for Research in Astronomy (SARA), USA}
\altaffiltext{3}{Department of Physics and Astronomy, East Tennessee State University, 
Johnson City, TN 37614, USA; hensong@mail.etsu.edu}

\begin{abstract}

New observations are presented of four evolved objects that display long, multi-year variations in their light curves.
These are interpreted as good evidence of their binary nature, with the modulation caused by the barycenter motion of the evolved star resulting in a periodic obscuration by a circumbinary disk.
Although protoplanetary nebulae (PPNe) commonly possess bipolar nebulae, which are thought to be shaped by a binary companion, there are very few PPNe in which a binary companion has been found.
Three of the objects in this study appear to be PPNe, IRAS 07253$-$2001, 08005$-$2356, and 17542$-$0603, with long periods of 5.2, 6.9, and 8.2 yrs, respectively.
The binary nature of IRAS 08005$-$2356 has recently been confirmed by a radial velocity study.
Two samples, one of PPNe and the other of post-AGB star candidates, are investigated for further evidence on how common is a long-period light curve variation.
Both samples suggest such light variations are not common.
The fourth object, IRAS 20056$+$1834 (QY Sge), is an obscured RV Tau variable of the RVb subclass, with a long period of 3.9 yrs and pulsation periods of 102.9 and 51.5 days.  The period of this object is seen to vary by 2$\%$.
Evidence is presented for a recent mass ejection in IRAS 17542$-$0603.\\
\end{abstract}

\section{Introduction}

Protoplanetary (or preplanetary) nebulae (PPNe) are a subclass of post-asymptotic giant branch (post-AGB) stars which have evolved from the main sequence through the red giant and asymptotic giant branch (AGB) phases, and which are now evolving horizontally across the Hertzsprung-Russell diagram toward higher temperatures.
PPNe are expected to become planetary nebulae (PNe) when their temperatures reach approximately 30,000 K and their ultraviolet photons can photo-ionize their surrounding nebula \citep{kwok00}.  This occurs in stars of low- and intermediate-mass (1$-$8 M$_{\sun}$), and the transition is expected to be short-lived (typically several thousand years).
The surrounding nebula is the remnant from the mass loss that occurred during the red giant and AGB phases, and thus they are surrounded by an expanding circumstellar envelope of gas and dust.  PPNe are distinguished from the general category of post-AGB objects by a compact nebula seen in scattered light at visible and near-infrared wavelengths and a cool ($\sim$200 K) dust shell detected in the mid-infrared.
A further classification made among post-AGB stars is to classify the infrared emission as due to a disk or shell, referring to the location of the dust. 
The disk sources have warm dust and are commonly found to be binary systems, with the binary thought to be responsible for the formation of the low-mass disk.
PPNe have cooler dust and are found to be shell sources \citep{vanwin19}

PNe are known to display a rich array of morphologies.  While some appear round (a ring in projection), many more have bipolar or "butterfly" morphologies or even more complex shapes. 
Given that they arise from spherical stars, this raises the question of the mechanism that shapes the ejected AGB envelope.
Strong evidence for binary companions has been found from systematic photometric studies of the central stars of PNe \citep[e.g.][]{jac21}, often followed up by spectroscopic monitoring that confirms this \citep[e.g.,][]{jones20}.  
A list of binary and suspected close binary central stars of PNe is maintained by David Jones\footnote{https://www.drdjones.net/bcspn}.
The role of the binaries in shaping the nebulae has been demonstrated by the agreement of the binary and nebular inclinations \citep{hil16}.
Almost all of the binary central stars found thus far in PNe have short periods ($<$10 days), and are thought to have evolved through a common envelope phase.  
However, only $\sim$20$\%$ of the PNe are found to have a close binary central star \citep{misz09}.  If binarity is important in shaping the nebulae, then there must be many cases which involve longer periods, which avoid the CE phase (P$\geq$1 yr).
Three PNe with multi-year periods have been found thus far through radial velocity measurements: BD+33 2642 (PN G052.7+50.7), NGC 1514, and LoTr 5, with orbital periods of 1105, 3306, and $\sim$2700 days, respectively \citep{vanwin14,jones17}.  These should have their antecedents in multi-year PPNe binaries and be even easier to detect in those of F$-$G spectral types because of their many and sharper lines.
Investigating even earlier in their evolution, a recent ALMA study of the shape of winds around a sample of AGB stars revealed bipolar morphologies, equatorial density enhancements, and complex arcs and spirals \citep{decin20}, all of which are seen in PNe and PPNe.  This led the authors to conclude that the same physical processes are at work in the AGB stars as in PNe, and that this is related to mass-loss rate and binarity.

While PNe have been observed for over 200 years, the study of PPNe is much more recent, blossoming following the detection of many cool infrared sources by the IRAS satellite and follow-up identifications beginning in the 1980s. 
Hubble Space Telescope observations showed that PPNe commonly display a bipolar morphology \citep{ueta00,su01,sah07,siod08}, indicating that the process that shapes the PNe begins early in this transitional phase or even in the preceding AGB phase.  
With this in mind, photometric and radial velocity monitoring studies were carried out on the brighter PPNe.  
However, evidence of binarity was not found.
Instead, these observations led to the discovery of variations due to pulsation, with periods in the range of 30 to 160 days \citep{hri10,ark10,ark11,hri13}.
Long-term radial velocity studies of seven of the brightest PPNe, carried out over a span of 25 years, found no definite cases and perhaps one suggested case of variations due to binarity \citep{hri11, hri17}.
This recently changed with the discovery of a binary companion in the PPNe IRAS 08005$-$2356 (V510 Pup), with a period of 2654 days (7.3 yrs) \citep{man21}.

In this study, we present photometric evidence of long, multi-year periodic variations in four post-AGB objects,
which we think are good evidence of their binary natures.
We begin with a brief description of the objects, discuss the observational data sets, and then describe the analyses of the observations.
The results are discussed in terms of a binary interpretation.
We conclude by reflecting on how common long periods might be among PPNe by examining some presently available samples of light curves.  
Binarity in these long-period cases can then be confirmed by radial velocity monitoring.
The goal is to test the binary hypothesis for the shaping of PPNe.

\section{Program Objects}

The program objects presented in this study were initially chosen for photometric monitoring based on their possession of a mid-infrared excess, indicating circumstellar dust.  
In two cases, early publicly-available light curves suggested a multi-year periodicity in the brightness. 
We list in Table~\ref{object_list} the four program objects, identifying them based on their IRAS (the main identifier used in this paper), 2MASS, and Gaia catalog numbers.
These are followed by their equatorial and galactic coordinates, apparent brightnesses, colors, and spectral types.  
They all lie within 10$\arcdeg$ of the galactic plane.  
Their spectral types are F and G, with supergiant luminosity classifications.
These stars are all redder than expected for their spectral types, which we attribute to a combination of interstellar and circumstellar material. 

All four of the objects display atomic emission lines in their visible line spectra.  
IRAS 08005$-$2356 shows many emission lines, some strong, with P Cygni profiles, which suggests some ongoing mass loss from the star \citep{sancon08}.
Strong lines of H$\alpha$ and H$\beta$ are seen in the spectrum of IRAS 17542$-$0603, along with many additional weak lines.  In contrast IRAS 07253$-$2001 shows only emission infilling at H$\alpha$, with H$\beta$ in absorption \cite{suarez06}.
IRAS 20056$+$1834 has a very unusual spectrum in that it displays very strong, broad Na {\it D} emission lines and also Ca~I 4226 $\AA$ emission lines superimposed on a G0 spectrum \citep{men88}.

\citet{men88} first called attention to the unusual spectral properties of IRAS 20056$+$1834.  The object was observed to vary in brightness, with near-infrared photometry indicating a  period of 50 days.  
They proposed a model in which the star is totally obscured by circumstellar dust and seen in reflected light.  Polarization studies support this model.
A high-resolution spectroscopic study by \citet{rao02} found a depletion pattern in the elemental abundances which showed a decreasing elemental abundance with increasing temperature of condensation, consistent with condensation in grains, 
It is similar to the pattern seen in RV Tau binaries \citep[e.g.,][]{maas02}.  
These properties led to the inclusion of IRAS 20056$+$1834 in a study of disks around post-AGB stars by \citet{deruy06}, in which the working hypothesis was that stable disks form due to the presence of a binary companion.
Variable radial velocities have been measured for this object  \citep{kloch07}, but the interpretation is not straightforward if they arise almost entirely from scattered light.
IRAS 08005$-$2356 is the only object in the study in which the nebula has been resolved; it has a clearly bipolar structure \citep{ueta00}.
Direct evidence for a binary central object has recently been found for this object based on its periodic radial velocity variability and photometric variability \citep{man21}.  
This present study documents the periodic variability in visible light.  It is one of the first PPNe to be found in a binary star system.
Both of these objects were included in a recent study of disks around evolved stars \citep{klus22}.

\section{Observations and Data Sets}
\label{obs}

We carried out observations at the Valparaiso University Observatory (VUO) and at the Southeastern Association for Research in Astronomy (SARA) Observatories for three of the four program objects.  We also made use of publicly-available data sets.  These are each described below and summarized in Table~\ref{data_sets}.

\subsection{VUO and SARA Observations}
\label{vuo-obs}

Observations were carried out  from the Valparaiso University Observatory (VUO) using the 0.4 m telescope for the two objects that can be observed from our northernly latitude, IRAS 17542$-$0603 and 20056$+$1834. 
Observations of IRAS 17542$-$0603 began in 1994 and those of IRAS 20056$+$1834 began in 2003.  Both were continued through 2018.
Two different camera systems were used.  The initial camera system did not have provisions for auto-guiding, and observations were restricted to a maximum of about seven minutes to preserve image quality.  These observations are referred to as VUO-old.  
This camera was replaced in 2008 with a better one, that included an auto-guiding system and a larger field of view.
This resulted in better image quality and more precise measurements.
We refer to these observations as VUO-new.
The Johnson {\it V} and Cousins {\it R}$_C$ filters were used.  Beginning in 1994 and for several seasons thereafter, the observations were mostly with the {\it V} filter, but starting in 2002, both filters were used each night.  
A defect in the {\it V} filter caused us to reject the {\it V} data from 2000 February through 2002 August, resulting in a gap in our {\it V} light curve of IRAS 17542$-$0603.  
The standardization of the VUO-old and VUO-new data from two the detector-filter sets did not include the second-order color term.  Since these program stars are red (({\it B$-$V})=0.7$-$1.4 mag), there may exist small, systematic offsets between the two data sets for each star, estimated to be on the order of a few hundredths of a magnitude.  
They are not evident in the light curves and are not expected to significantly affect the results.

Observations were also carried out for two of the three southern hemisphere objects using two of the telescopes operated by the Southeastern Association for Research in Astronomy  \citep[SARA;][]{keel17}, of which Valparaiso University is a member.  These were the 0.9 m telescope at Kitt Peak, AZ (SARA-KP) and the 0.6 m Lowell Telescope at Cerro Tololo, Chile (SARA-CT).  
Observations were carried out from 2009 through 2018 using the {\it V} and {\it R}$_C$ filters. 

The data were all reduced using standard image reduction procedures of cosmic ray removal, bias/dark subtraction, flat fielding, and aperture photometry, using standard procedures in IRAF\footnote{IRAF is distributed by the National Optical Astronomical Observatory, operated by the Association for Universities for Research in Astronomy, Inc., under contract with the National Science Foundation.}.
We used apertures of $\sim$5$\farcs$5 radius.
Transformations to the standard system were made based on observations of Landolt stars  \citep{land83, land92}, using linear color coefficients.
Our observed standardized magnitudes and colors of the program stars, obtained on the best photometric nights, are listed in Table~\ref{std_ppn}.
We carried out differential photometry, using three comparison stars for each source to check their internal consistency.  These are identified in Table~\ref{std_comp}, where we have also listed their standard magnitudes and colors. 
These comparison stars were found to be constant at the level of $\pm$0.01 to $\pm$0.02 
mag over the interval of observations. 

The camera systems at these two SARA telescopes were changed several times during our observing interval. 
As we found in two previous studies, there appear to be systematic photometric offsets for each star between the camera systems.  We attribute these to the neglect of second-order color terms in the standardization of the observations for these reddened stars.  
These offset values were determined empirically and are listed in the Appendix, Table~\ref{offsets}, where they are discussed in more detail.
 
\begin{figure}\figurenum{1}\epsscale{1.5} 
\plotone{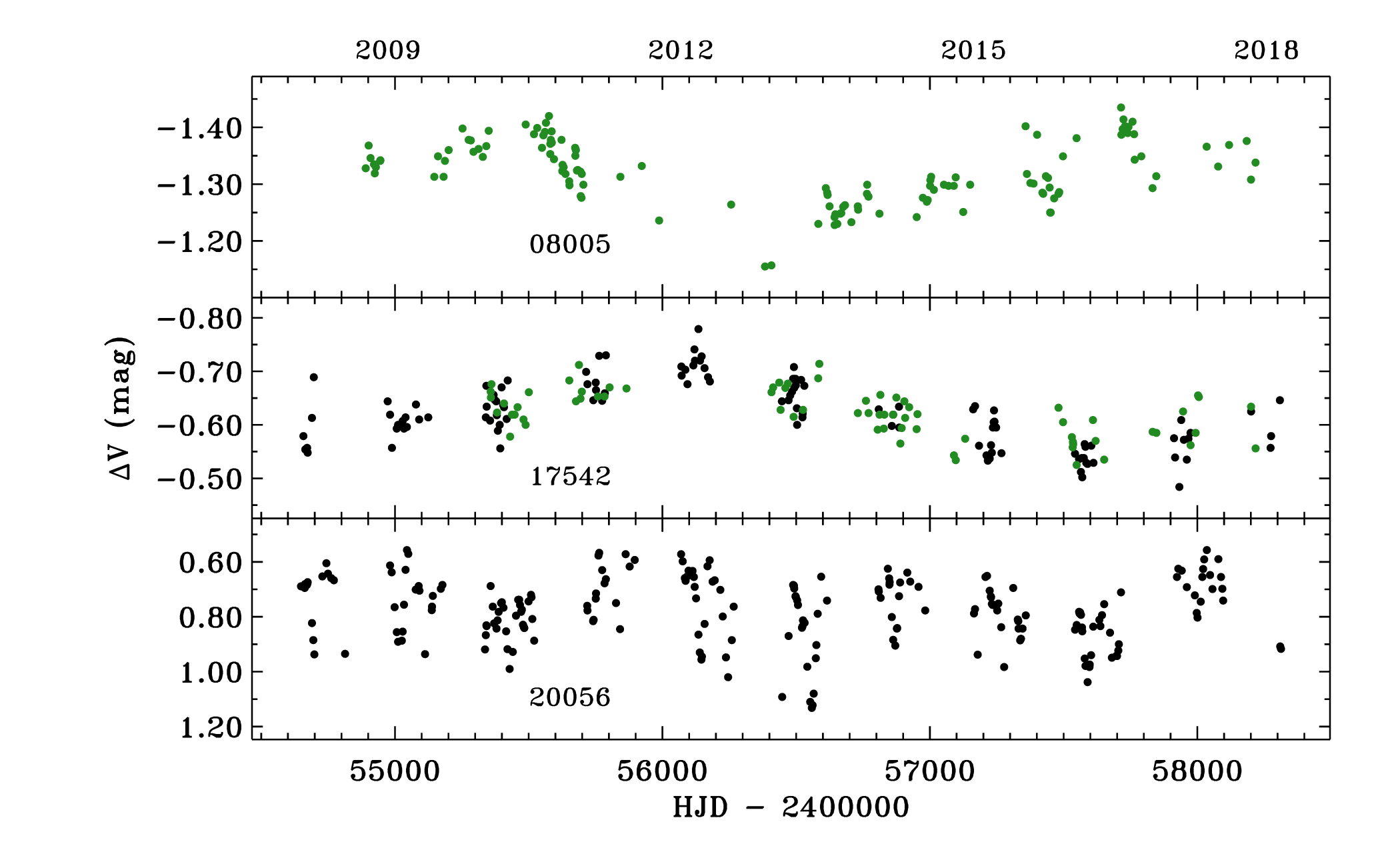}
\caption{Our new {\it V} light curves from 2008$-$2018.  The VUO-new data are shown in black and the SARA data in green.
\label{LC-VU-SARA}}
\epsscale{1.0}
\end{figure}

The {\it V} light curves from 2008$-$2018 for the three objects observed with the VUO and SARA telescopes are displayed in Figure~\ref{LC-VU-SARA}. 
The objects all vary in brightness, with ranges in a season of up to 0.14 mag for IRAS 08005$-$2356 and 17542$-$0603 and 0.48 mag for IRAS 20056$+$1834.
The overall ranges are 0.28, 0.24, and 0.58 mag for IRAS 08005$-$2356, 17542$-$0603, and 20056$+$1834, respectively, over the 10$-$11 year interval of observations.
The individual observations for each object are accessible electronically (see Appendix, Table~\ref{std_mags}).

\subsection{ASAS-SN Data}
\label{asas-sn-data}

These four objects were all observed as part of the All-Sky Automated Survey for Supernovae \citep[ASAS-SN;][]{koch17} monitoring program. 
These observations were made through {\it V} and {\it g} (the Sloan Digital Sky Survey {\it g} filter, $\lambda$$_{\rm eff}$=477 nm) filters 
mounted on cameras with 14 cm aperture telephoto lenses and CCD detectors. 
Aperture photometry was used to determine the magnitudes, using a relatively large radius of 16$\arcsec$.   The data were standardized and are publicly available\footnote{https://asas-sn.osu.edu}.
Typically, three consecutive, 90 sec dithered observations were made on a night.  We combined these to determine an average magnitude, first eliminating data of inferior quality.  
The statistical uncertainties in the ASAS-SN measurements of these objects are approximately $\pm$0.005$-$0.009 mag.
Multiple cameras were used in acquiring the data, and we found that there existed small, systematic offsets in magnitude between the measurements of the stars made with the different cameras.
These offsets were empirically determined for each star by inter-comparing measurements made on the same or adjacent nights with the different camera systems, and the offsets were applied to the data.
This is discussed more fully in a previous paper in this series \citep{hri21}.

\begin{figure}\figurenum{2}\epsscale{1.5} 
\plotone{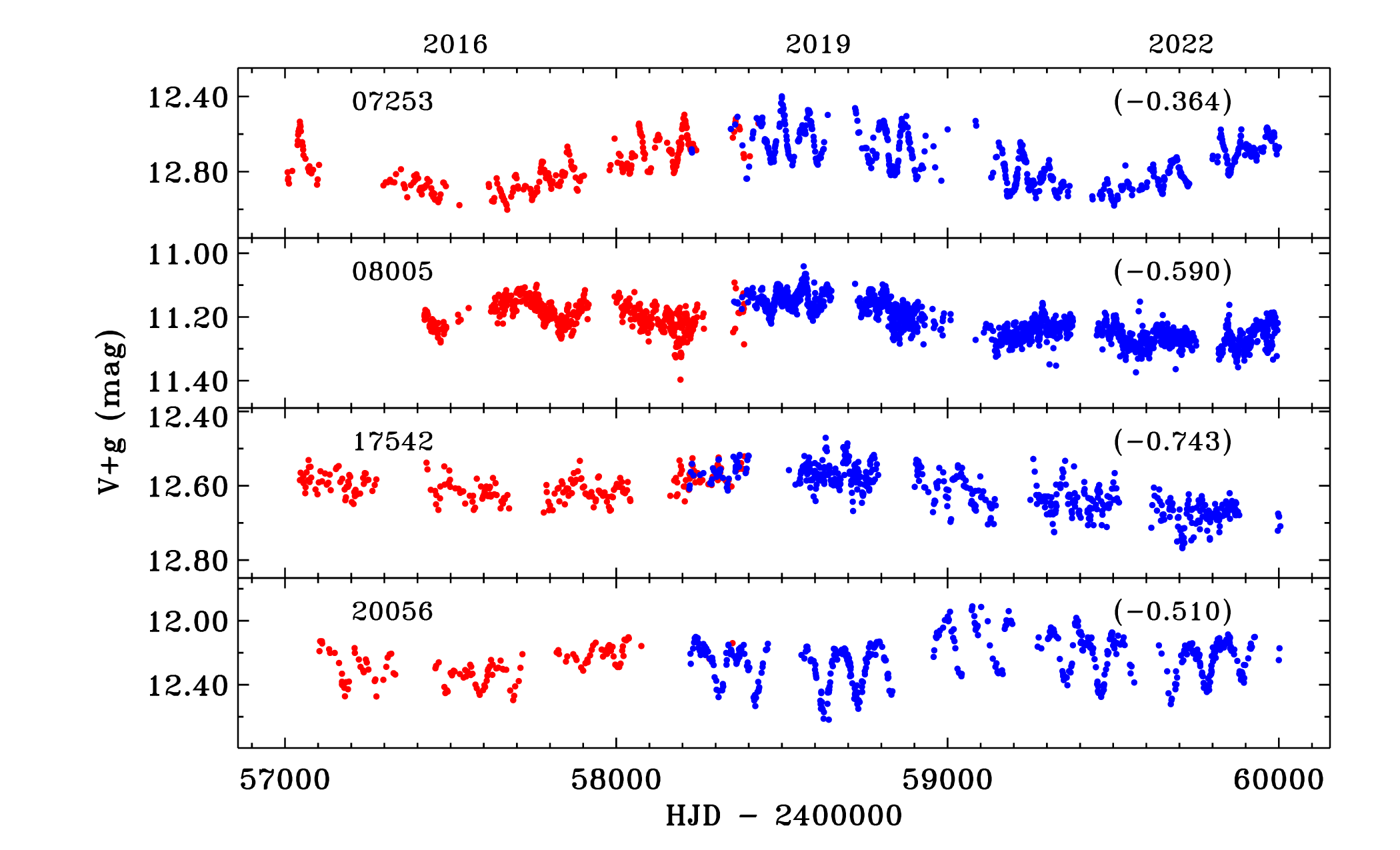}
\caption{The ASAS-SN {\it V} light curves (red) from 2015 to 2018 and {\it g} light curves (blue) from 2018 to 2023, with the later offset to the level of the {\it V} light curve for the interval of overlap.  The amount of the offset in units of mag is listed in parentheses. 
Note that the light curves of IRAS 17542$-$0603 are contaminated by $\sim$0.2 mag by a nearby star in the large ASAS-SN photometric aperture.    
\label{asas-sn}}
\epsscale{1.0}
\end{figure}

The {\it V} observations of the stars in this study began in winter 2014$-$2015 for three of the objects and winter 2016 for the fourth object, and they continued to the latter half of 2018.  The {\it g} observations began in Apr or Aug 2018, and we used data acquired through Feb or Mar 2023.  
The observing dates for each of the data sets are listed in Table~\ref{data_sets}. 
This resulted in three or four seasons of {\it V} observations and five of {\it g}, with some slight temporal overlap in observations with the two filters.
The light curves are displayed in Figure~\ref{asas-sn}.
In order to follow the longer-term variability in the data, we have shifted the {\it g} data by an empirically-determined offset derived from the interval of overlap, so that it agrees with the level of the {\it V} observations.
The multi-year variations in the data can be seen in all four of the program objects, along with clear, shorter-term variations in IRAS 07253$-$2001 and 20056$+$1834, especially evident in the {\it g} filter.
Note that previous studies of PPNe have shown that pulsation amplitudes are larger at shorter wavelengths \citep{hri13,hri18}.

\subsection{ASAS-3 Data}
\label{asas-data}

Earlier photometric survey data from 2000$-$2009 are available in the  {\it V} filter for each of these four objects from the All Sky Automated Survey, version 3 (ASAS-3)\footnote{http://www.astrouw.edu.pl/asas/}. 
This survey used a camera with a 20 cm, f/2,8 telephoto lens and a CCD detector in which the pixels were 15$\arcsec$ in size.  
Although less precise than our pointed observations or the ASAS-SN survey data, they are useful for investigating long-term trends in the brightness of the objects.
To reduce contamination by nearby stars, we used only the data measured with the smallest aperture (15$\arcsec$ in radius).
Something appears to be amiss in the ASAS-3 {\it V} measurements of IRAS 20056$+$1834, as they are $\sim$0.8 mag brighter than what we measured with a smaller aperture at the VUO, even during the same time interval.  The nearest bright star is the comparison star C1, which is 0.65 mag brighter and located 24$\arcsec$ SE of the star.  While one might expect there to be some contamination, this seems rather extreme, and so we decided not to use the ASAS-3 data for this star. 
We selected only the best quality data (classified as Grade=A) for the four objects and eliminated measurements with $\sigma$ $\ge$ 0.065 mag.
A few very discrepant points were removed from the datasets.  
Average uncertainties in the data we used are $\pm$0.040$-$0.045 mag.

\section{Light Curves and Variability Study}

\subsection{Analysis Strategy}
\label{analysis}

Our strategy was to begin the study of each star with an analysis of its light curves in the individual data sets (VUO-new, VUO-old, SARA, ASAS-SN) in the various filters, to see what consistency was found among them.  We then combined the various {\it V} data sets into a combined light curve for each star, which gave us a baseline of 15 to 19 years.
To combine the standardized differential VUO and SARA {\it V} data with the standardized ASAS-SN and ASAS-3 data, an offset was added to the differential data.  
We determined the offset empirically by comparing the difference between the VUO-new and SARA data compared to the ASAS-SN when observed on the same or adjacent nights.  

We carried out a periodogram analysis of the observations using the program PERIOD04 \citep{lenz05}, which uses a Fourier analysis to search for the most likely periods in the frequency domain.  The program also allows one to search for multiple periods simultaneously, which are useful if there is both shorter-term pulsations (30$-$160 days) and longer-term (multi-year) variations due to binarity.  Following typical practice, we considered a peak in the power spectrum to be significant if its signal-to-noise ratio (S/N) was greater than 4.0 \citep{bre93}.  In all cases, the values were much higher.

\subsection{IRAS 08005$-$2356} 

We were aware that the light curve of IRAS 08005$-$2356 possessed a long-period, cyclical (sinusoidal) variation based on the early ASAS-3 data.
This motivated us to carry out SARA observations of IRAS 08005$-$2356.
Thus we have several data sets to analyze, both individually and in combination.
The SARA data (2009$-$2018) show evidence of a long-period cyclical behavior in both the {\it V} (see Fig.~\ref{LC-VU-SARA}) and the {\it R}$_C$ light curves.  
Analyses of these result in periods of 7.1$\pm$0.2 yr (2580$\pm$80 days) and 7.3$\pm$0.2 yr (2660$\pm$90 days), respectively. 
The period for the less precise ASAS-3 data (2000$-$2009) is shorter, 6.3$\pm$0.1 yr, with a larger amplitude.
The ASAS-SN data show a cyclical variation similar to the SARA data, as shown in Figure~\ref{asas-sn}.
The ASAS-SN {\it g} light curve covers only five seasons and yields a period of 7.5$\pm$0.3 yr.
The ASAS-SN {\it V} data agree well with the SARA data. 
When combined, they yield a period of 6.6$\pm$0.1 yr. 
Investigating the period in these different sets and combinations of sets thus yields a range of periods from 6.3 to 7.5 yr.  

\begin{figure}\figurenum{3}\epsscale{1.5} 
\plotone{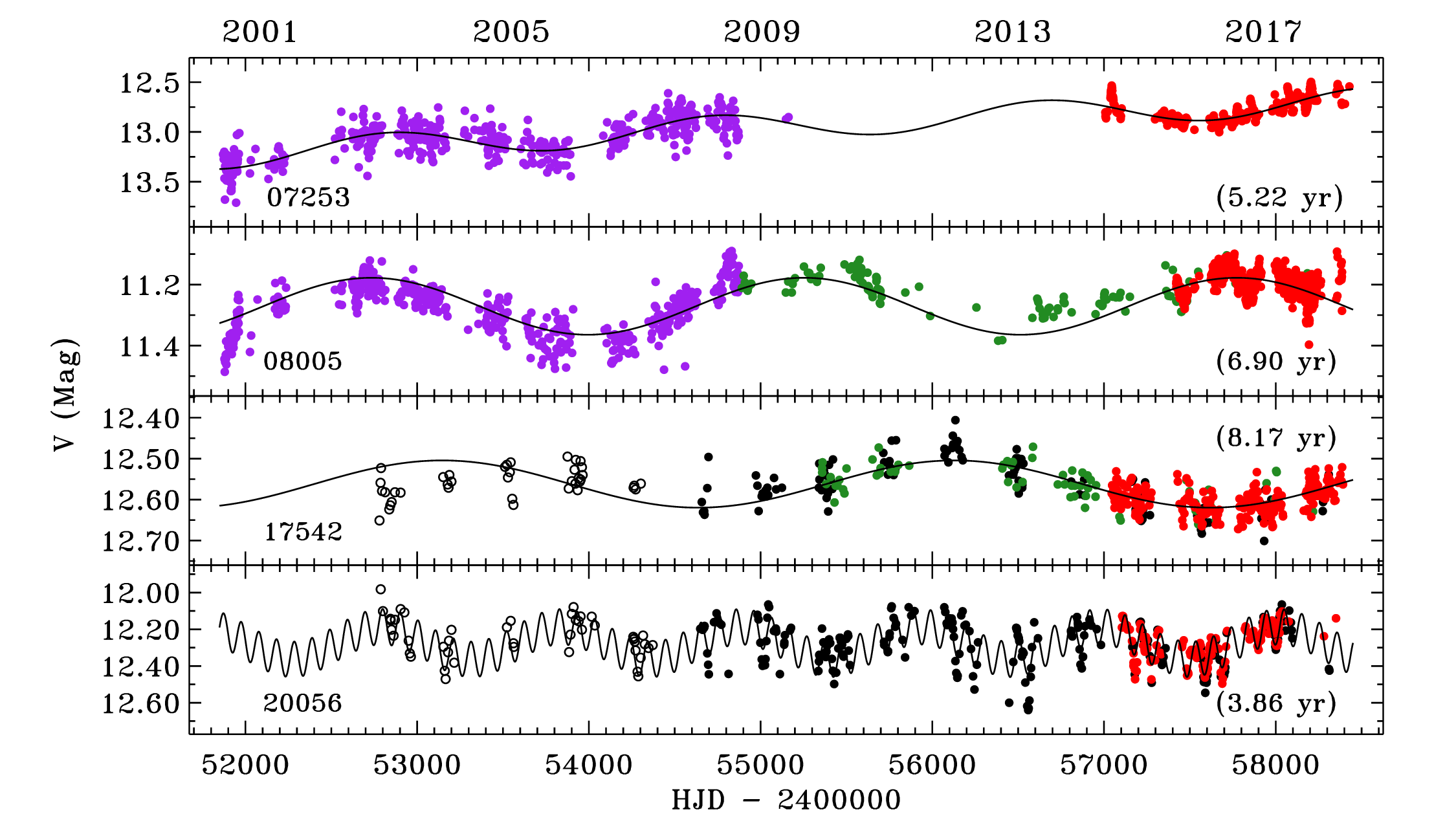}
\caption{The combined {\it V} light curve of the program objects, based on ASAS-3 (purple), VUO-old (white), VUO-new (black), SARA (green), and ASAS-SN (red) data.
The solid curve represents the fit to the periodogram solution as listed in Table~\ref{periods}.  The short period is also included for IRAS 20056$+$1834 but not for IRAS 07253$-$2001 (see Fig.~\ref{07253}). 
\label{lc-fit}}
\epsscale{1.0}
\end{figure}

Combining all of the {\it V} data results in a period of 6.9$\pm$0.1 yr, where we have given the SARA and ASAS-SN data twice the weight of the less-precise ASAS-3 data.  The semi-amplitude is 0.09 mag.
The fit of this solution to the light curve is shown in Figure~\ref{lc-fit}.
The parameters of the periodogram fits are listed in Table~\ref{periods}, both of the long-term (multi-year) and, where applicable, the shorter-term (pulsation) variations.
While the solution agrees well with the general cyclical variations in the light curve, there are obvious seasonal deviations, which we suggest are due to the variations in the dust opacity of the circumstellar material.  
In addition, there is a large range in the brightness variations within a season, up to 0.20 mag in the ASAS-SN data and even larger in the less-precise ASAS-3 data.  These we attribute to instabilities in the star, although we do not find a stable period that we would attribute to pulsation.  The 2019 ASAS-SN {\it g} data appear to show a period of 70 days, but that is not found in any of the other seasons or overall data. 

\subsection{IRAS 17542$-$0603}.

IRAS 17542$-$0603 was chosen for observations based upon its excess emission in the IRAS 25 $\mu$m bandpass.  
Observations began in 1994 with the {\it V} filter, with a few {\it R}$_C$ observations beginning the next year.  Starting in 2002, observations were made regularly using both filters and this continued through 2018.   
SARA {\it V} and {\it R}$_C$ observations began in 2010 and continued through 2018, except for 2012.  
In Figure~\ref{LC-VU-SARA} is shown the combined VUO-new and SARA {\it V} light curve from 2008$-$2018.
It is seen to vary in an approximately sinusoidal pattern over the 11 seasons of observing, increasing in brightness by $\sim$0.12 mag over the first five seasons, reaching a maximum in 2012, then decreasing in brightness by $\sim$0.17 mag over the next 4 seasons, and finally increasing again in the last two.  This suggests a period of $\sim$9 years.
The average peak-to-peak variation in a season is 0.10 mag in {\it V} and {\it R}$_C$.
The combined ASAS-SN {\it V} and {\it g} data also show a cyclical pattern (see Fig.~\ref{asas-sn}).
The four seasons of {\it V} data show a minimum in the 2016 and 2017 seasons and then the overall light curve shows a maximum in the 2018 and 2019 seasons, with the {\it g} data then showing a brightness decrease of $\sim$0.11 mag to a lower minimum value in the blue light curve.   

\begin{figure}\figurenum{4}\epsscale{1.5} 
\plotone{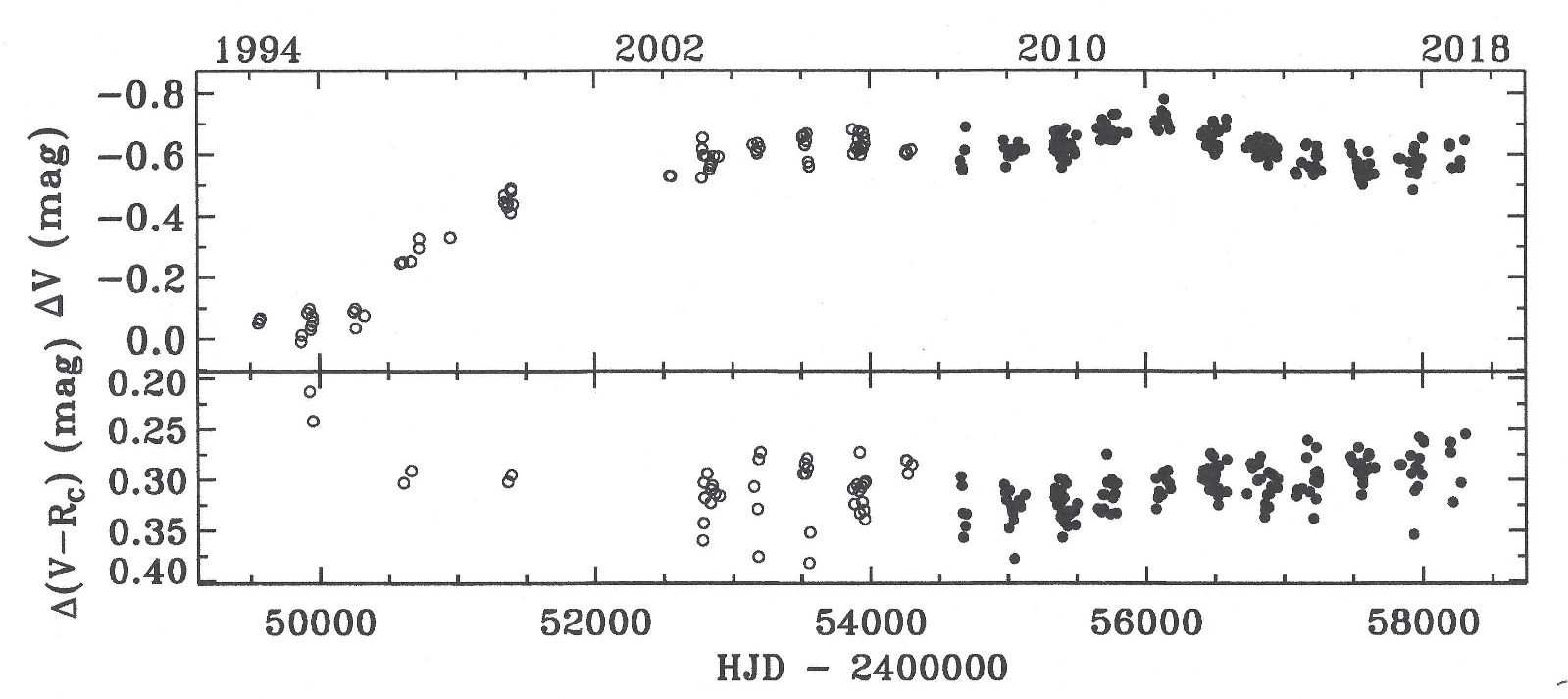}
\caption{The light and color curves of IRAS 17542$-$0603 from 1994$-$2018, based on VUO-old (open circle) and VUO-new (filled circle) data.  
(top) The {\it V} light curve.
(bottom) The color curve.
\label{17542}}
\epsscale{1.0}
\end{figure}

We also examined the earlier, less precise VUO-old data from 1995$-$2007, as shown in Figure~\ref{17542}, along with the color change.
One can see that the light curve varies over a much larger range of 0.60 mag ({\it V}), 
increasing in brightness from $\sim$1995 
to 2005$-$2006 and then smoothly joining with the newer data at what appears to be a local minimum at $\sim$2007, in addition to the one in 2016$-$2017.  
The {\it R}$_C$ light curve, which includes data from 2001 and 2002, suggests that there may be another local minima in the light curve around 2001$-$2002.  

The ({\it V$-$R}$_C$) color index also changes over a rather large range.  
The color is significantly less red during the deep minimum in 1995 and then increases in color index by $\sim$0.08 mag over the next few years.  
Thereafter, the color variation is less, with the VUO-old data displaying a larger seasonal range, which is most likely the result of the lower precision of the VUO-old data.
This change between the color in 1995 and later years, when the system in brighter and redder, is supported by the standardized observations that we made of the system in 1994 at Kitt Peak National Observatory, which are listed in Table~\ref{std_ppn}.
The color index is seen to decrease approximately monotonically by about $-$0.05 mag over the 11 seasons of VUO-new and SARA observations (2008$-$2018), becoming less red.  

We began by analyzing the different data sets separately and in combinations.
Analyzing the VUO-new, the VUO-new and SARA, the VUO-new and ASAS-SN, and the VUO-new, SARA, and ASAS-SN {\it V} light curve data, and found periods in the range of 8.4 to 8.9$\pm$0.3 yr.  
From the VUO-new and the VUO-new and SARA {\it R}$_C$ light curves we found periods of 9.7 to 10.1$\pm$0.3 yrs.  
To include the earlier VUO-old data in the light curve analysis, we need to make some adjustments to the data or add extra terms to accommodate the pronounced faintness of the data before 2002.  We chose instead to simply use the data from 2003 onward, which eliminated the earlier, fainter data, and this results in a light curve which appears sinusoidal around a median value.  
We examined the ASAS-3 data from 2000$-$2009 and compared it with our VUO-old data.  The data showed good agreement, but the ASAS-3 data were much less precise and so we did not use them in the period analysis.
We note that there is a nearby star (2MASS 17565584$-$0604018) $\sim$9$\arcsec$ NW of the object and somewhat fainter.  It contributes to the light of the program star in the larger ASAS-SN and ASAS-3 apertures, increasing it by $\sim$0.2 mag ({\it V}).

For our final solution, we used all of the {\it V} data from 2003 to 2018, giving the later three data sets (VUO-new, SARA, ASAS-SN) twice the weight of the less precise VUO-old data. 
This results in a period of 8.2$\pm$0.1 yr and a semi-amplitude of 0.06 mag.  The results are listed in Table~\ref{periods} and the fit to the light curve from 2003 to 2018 is shown in Figure~\ref{lc-fit}.
This fits the observations reasonably well, except for the 2003-2004 data, which may still be affected by the brightness decrease seen in the earlier years.  
We searched for evidence of a shorter, pulsational period, particularly in the high density ASAS-SN data, but none was found.

We still have the mystery of the increase of brightness by $\sim$0.5 mag from 1994 to 2002.   
Without earlier data, it is unclear if the object itself increased in brightness or if its brightness was instead recovering to its former level following an obscuring event that occurred around 1994.  
An argument for the latter is the fact that the object appeared bluer when fainter.  This could be explained by a mass loss event which led to the formation of dust around the star, which dimmed the light but also scattered more blue light into our line of site.
As this material expanded away from the star and dispersed, the star appeared to brighten and the scattered bluer light decreased and became less significant.
Variable dust-obscuring events have been seen in several PNe \citep[and references therein]{paun23}. 

\subsection{IRAS 20056$+$1834} 

IRAS 20056$+$1834 (QY Sge) is a known variable ascribed to be of the RV Tau type. 
Observations of it were made at the VUO from 2003 through 2018 using the {\it V} and {\it R}$_C$ filters.  
In addition to the short-term cyclical behavior, they also show to a wave-like pattern in the overall light levels (Fig.~\ref{LC-VU-SARA}).  Analyses revealed the identical three periods in both the {\it V} and {\it R}$_C$ data sets: {\it P}$_{\rm long}$ = 3.85$\pm$0.05 yr, {\it P}$_1$ = 102.9$\pm$0.1 days, and {\it P}$_2$ = 51.5$\pm$0.1 days.
The amplitude of {\it P}$_1$ is larger in the shorter {\it V} bandpass, while the amplitudes are approximately the same for each of {\it P}$_2$ and {\it P}$_{\rm long}$ in the {\it V} and {\it R}$_C$ bandpasses.  
The ASAS-SN data also show both the shorter-term and the multi-year variations.  
The ASAS-SN {\it V} light curve covers only three years, not long enough to give a reliable long period but does show similar short periods of 102.5$\pm$0.3 days and 51.5$\pm$0.1 days.
Combining the {\it V} light curves, giving the more precise VUO-new data twice the weight of the others, results in {\it P}$_{\rm long}$ = 3.85$\pm$0.05 yr, {\it P}$_1$ = 102.9$\pm$0.1 days, and {\it P}$_2$ = 51.5$\pm$0.1 days.  The results of this analysis are listed in Table~\ref{periods} and shown in Figure~\ref{lc-fit}.
There is evidence for a small amount of contamination in the ASAS-SN data by the brighter star $\sim$24$\arcsec$ SE of our program star.  
The color varies with the pulsation phase, and an analysis of the ({\it V$-$R}$_C$) data gives the same values of {\it P}$_1$ = 102.9$\pm$0.1 days and {\it P}$_2$ = 51.5$\pm$0.1 days.  

We also analyzed the large data set of ASAS-SN {\it g} data taken over five seasons.  
The shorter-term variations have larger amplitudes in the shorter {\it g} filter, and surprisingly, they have slightly longer (by 2$\%$) periods of 104.8$\pm$0.1 and 52.4$\pm$0.1 days during this time interval (2018$-$2023).  This is surprising, as we had not encountered a difference in period at different bandpasses in our previous studies of post-AGB stars \citep{hri13,hri17}, in which multiple bandpasses were measured contemporaneously.  This suggests that a change in period occurred over the interval of a few years.  
We investigate this further by analyzing year by year the data sets that covered two deeper minima in a season reasonably well.  
These included all five years of the ASAS-SN {\it g} data, all three years in which we had both ASAS-SN and VUO-new {\it V} data, and two additional VUO-new {\it V} datasets (2009, 2012).
The periods for the {\it g} light curves, between 2018 and 2022, ranged from 104.5 to 106.9 days, with an average of 105.3 days; the {\it V} light curves between 2015 and 2017 had a range of 98.2 to 102.3 days, with an average period of 100.2 days; the two additional VUO-new {\it V} data sets (2009, 2012), had values of 100.3 and 102.7 days, with an average of 101.5 days.  
Thus, these yearly data sets consistently show values in accord with the overall fits in these time intervals and bandpasses and support the claim of a period change in 2018. 
A similar period value of 104.7 days, based on recent Zwicky Transient Facility measurements \citep{chen20}, which began in 2018, confirms this change to a longer period.
Period changes are well known to occur in RV Tau variables \citep{per97}.

\subsection{IRAS 07253$-$2001} 

\begin{figure}\figurenum{5}\epsscale{1.5} 
\plotone{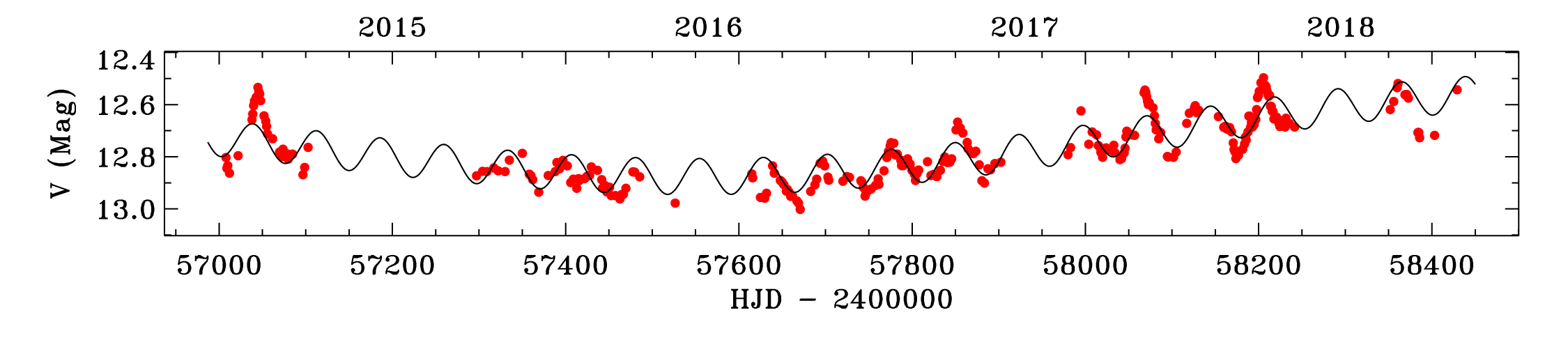}
\caption{IRAS 07253$-$2001: the fit of the combined solution to the ASAS-SN {\it V} data, including the short period of 73.6 days.
\label{07253}}
\epsscale{1.0}
\end{figure}

IRAS 07253$-$2001 was the one object in this study which we did not monitor photometrically.  We had available only ASAS-SN {\it V} and {\it g} and ASAS-3 {\it V} data.  All show evidence of multi-year cyclical variations.  
The ASAS-SN data also display variations on a shorter time scale, as shown in Figure~\ref{asas-sn}.   
These appear to decrease in amplitude during the minima of the long period variations.
We analyzed the different data sets separately and in combinations.  They each possess a long period in the range of 4.3$-$5.0 years and a short period of 73 days. 
Since the ASAS-SN {\it V} and {\it g} data only cover four and five seasons, respectively, which is about the length of the long period, we also combined them into a longer nine year interval by empirically adjusting the {\it g} measurements to the level of the {\it V} measurements.  The precision in this offset is not high, $\pm$0.02 mag, since the temporal overlap is small, but is appears to be sufficient for our purposes.  This results in a long period of 5.0$\pm$0.1 yrs 
and a short period of 73.0$\pm$0.1 days.  The analyses of the ASAS-SN data also suggests a second short period of 69 days, although this may be the result of an attempt to fit the variable amplitudes. 
The early ASAS-3 {\it V} light curve shows evidence of a general increase in brightness of $\sim$0.02 mag over the interval 2000$-$2009.
The analysis of the ASAS-3 data results in periods of 5.0$\pm$0.1 yr 
and 73.8$\pm$0.1 days, 
in addition to a very long period to account for the gradual increase in brightness. 
Finally, we combined all of the {\it V} data, with the ASAS-SN data given twice the weight of the less precise ASAS-3 data.  This resulted in a long period of 5.2$\pm$0.1 yr and a shorter period of 73.6$\pm$0.1 days\footnote{We note that, in a recent, independent study of IRAS 07253$-$2001, \citet{ikon23} determined a long period of 1810 days (5.0 yrs), a dominant pulsational period of 73 days, and additional weaker periods of 68 and 70 days.  These are based on an analysis of the ASAS-SN data and their own more recent observations.  The period results of the two studies are in good agreement.}.
These results are listed in Table~\ref{periods}.  
In Figure~\ref{lc-fit} is shown the good fit to the overall {\it V} light curve by the long period with the gradual increase in brightness included.  The fit to the more precise ASAS-SN {\it V} data, including the shorter, pulsation period, is shown in Figure~\ref{07253}.

\section{DISCUSSION}

\subsection{Spectral Energy Distributions and Post-AGB Nature}

These four objects were chosen for photometric study because each one possessed an infrared excess that made it a candidate post-AGB star.
An examination of the spectral energy distribution (SED) of each appears to confirm that evolutionary classification.
Each one displays a peak in the red end of the visible spectrum, 0.8$-$0.9 $\mu$m, and then one or two additional peaks between 1 and 100 $\mu$m.
The second peak for IRAS 20056$+$1834 is seen at 6$-$12 $\mu$m.
IRAS 07253$-$2001 possesses a broad second peak reaching maximum between 10 and 25 $\mu$m, or possible two blended peaks at 10 and $\sim$25 $\mu$m.
IRAS 08005$-$2356 possesses two additional peaks, one at 6 $\mu$m and another at $\sim$25 $\mu$m; the SED of IRAS 17542$-$0603 is somewhat similar, but with peaks at 4 and at $\sim$25 $\mu$m.
In all four cases, the flux received in the infrared is larger than that in the visible, and for IRAS 07253$-$2001 and 20056$+$1834, it is larger by approximately a factor of ten.
This is not surprising for IRAS 20056$+$1834, since it is thought to be completely obscured and seen in the visible only by scattered light.  
The SEDs of IRAS 08005$-$2356 and 20056$+$1834 have been displayed by \citet{klus22}.  
An detailed spectropolarimetric modeling study of IRAS 08005$-$2356 was carried out by \citet{open05}. 
They fit the SED with an inner region with strong latitudinal dependance shell (effectively a thick disk) of temperature 1065 K, an outer shell of inner temperature 170 K, with the disk at an inclination of 60$\pm$5$^o$.  Their model also fits the images well.
\citet{ram17} have included IRAS 08005$-$2356 and 17542$-$0603 in their infrared photometric and spectroscopic study of a sample post-AGB stars.  They fit the infrared observations of IRAS 08005$-$2356 with a single shell of temperature 440 K and IRAS 17542$-$0603 with two shells of temperature 1000 K and 500 K.  
Thus, IRAS 20056$+$1834 appears to have warm dust and the other three have both warm and cool dust.

It is common to ascribe the $\sim$25 $\mu$m peak to the emission from the cool (T$\sim$200 K) dust in the distended, expanding circumstellar envelopes of PPNe \citep{hri89}.
The 6$-$10 $\mu$m peak, in contrast, is ascribed to warm dust from a bound circumstellar disk with a range of temperatures.
These disk sources are commonly found to be post-AGB binaries, and they have been studied extensively by Van Winckel and collaborators \citep{vanwin95,oom18,klus22}. 
Binarity has particularly has been established in RV Tau variables with these dusty disks \citep{man17}.
It is unusual to find the near-infrared emission from warm dust in PPNe.

The recent availability of good parallaxes for so many stars in our galaxy allows one to use these SEDs to determine their luminosities, with account taken of interstellar and circumstellar extinction and, in some cases, the geometry of the circumstellar matter.  
 Recently \citet{oud22} published a very extensive study of the luminosities of post-AGB stars.  Combining integrated fluxes for the stars with their distances, they determined luminosities of 185 objects and were able to place 134 with spectral types on an HR diagram.  The distances were based on Gaia DR3 data using the results of \citet{bai21} for geometrical distances, and the integrated fluxes were from the study of \citet{vic15}.  
 Three of our targets are included in this study, and the distances and resulting luminosities for two of them are very different from what one would expect for post-AGB stars.  The distances and luminosities found are as follows: IRAS 07253$-$2001 $-$ 0.45 kpc, 26 {\it L}$_{\sun}$: IRAS 08005$-$2356 $-$ 9.2 kpc, 54,000 {\it L}$_{\sun}$; and IRAS 17542$-$0603 $-$ 5.9 kpc, 4700 {\it L}$_{\sun}$. Only for IRAS 17542$-$0603 does the luminosity fall within the expected range of 1600$-$20,000 {\it L}$_{\sun}$ as calculated in the models of \citet{milb16}.  
 However, as \citet{oud22} point out, some of the parallaxes have large uncertainties, and these are indicated by large values for the renormalized unit weight error (RUWE).  They cite a typical value for the RUWE of $<$ 1.4, which is normally an indication of a good quality.  For the three above objects, the RUWE are 43, 6, and 6, respectively.  \citet{oud22} state that the RUWE is sensitive to the presence of a binary companion and extended emission, 
 and in our case, we would expect these very factors to degrade the parallax measurements for these three stars.  
 IRAS 20056$+$1834 was not included in their study; its Gaia DR3 distance is 3.3 kpc \citep{bai21} and the RUWE is 14.
Thus, we suspect that the distance determinations for all four of these objects are compromised by their likely binary motions. 
In fact, the large RUWE values and the apparently discordant luminosities for two of those in the \citet{oud22} study may be indirect evidence for their binary natures.\footnote{\citet{partha23} also made a similar argument for objects with discordant luminosities in the study of post-AGB stars by \citet{kam22}.} 

Given that the distances are likely compromised for these four objects, and are not reliable as a means to determine luminosity, 
one can investigate the stellar and circumstellar chemistry to see if it can provide some additional information to support the post-AGB the nature of these objects.
IRAS 17542$-$0603 possesses a mixed carbon and oxygen circumstellar chemistry, displaying both UIR (C-rich) features and the 9.7 $\mu$m silicate (O-rich) feature in its mid-infrared spectrum \citep{cer09}.  This is seen in only a small number of evolved objects.  IRAS 08005$-$2356 also displays a mixed carbon and oxygen spectrum, but it is unique in the lines that it displays, both C$_2$ and  C$_3$ absorption lines in its visible spectrum \citep{slijk91,hri95} and molecular-line OH \citep{lik89} but not CO lines in its circumstellar envelope.  It has earlier been suggested to be an evolved object in transition from being oxygen-rich to carbon-rich \citep{hri95}.
IRAS 07253$-$2001 appears to be O-rich based on the presence of 9.7 $\mu$m silicate feature in absorption in N-band photometry \citep{blom93}, although it has not been detected in OH \citep{hek91}, SiO, or H2O radio emission lines \citep{yoon14}.
The chemistry is consistent with the objects indeed being post-AGB.

\subsection{IRAS 20056$+$1834 as an RV Tauri Variable} 

\begin{figure}\figurenum{6}\epsscale{1.5} 
\plotone{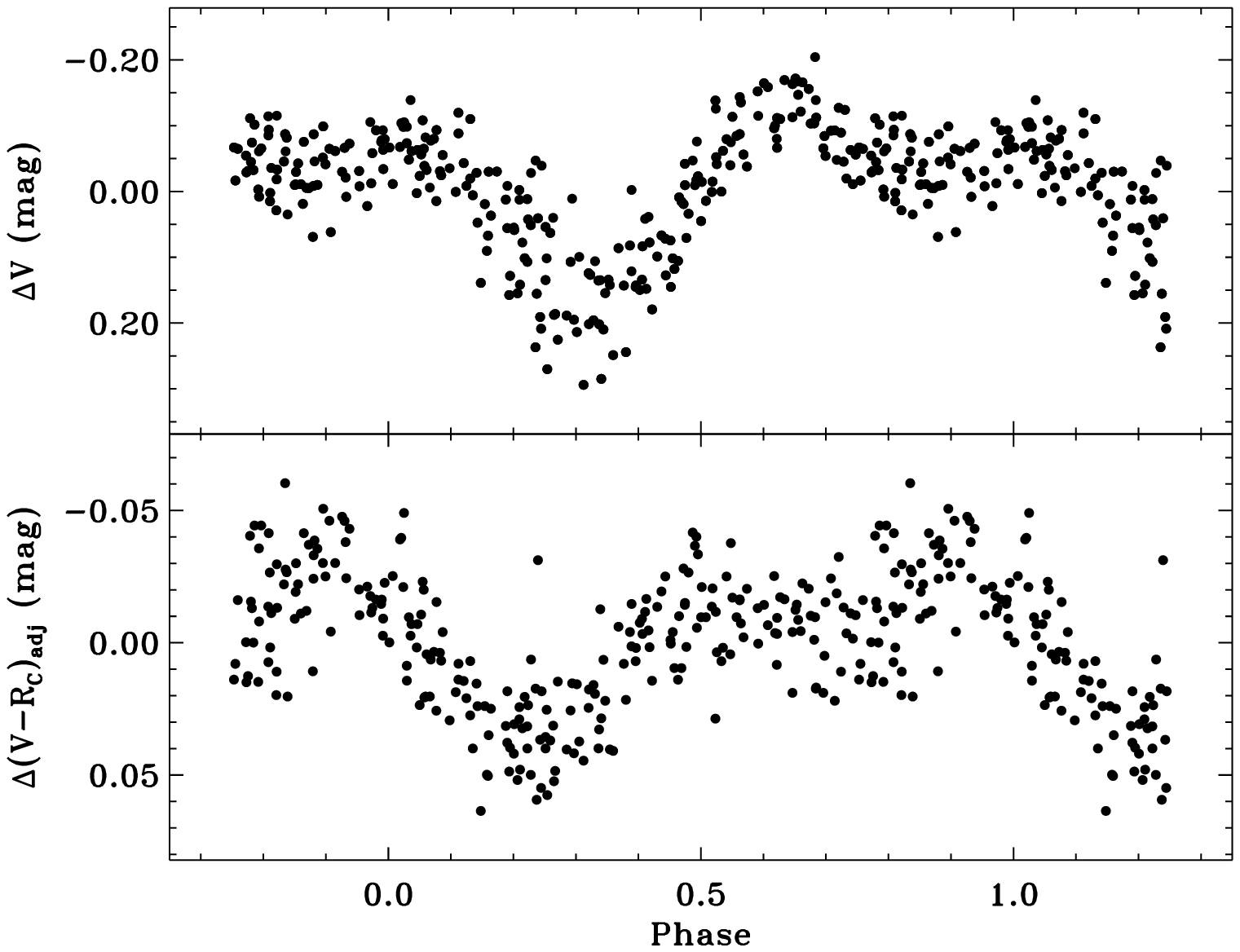}
\caption{IRAS 20056$+$1834: (top) Phased {\it V} light curve, with the long-period variation removed.  
(bottom): Phased ({\it V$-$R}$_C$) color curve, using the same epoch and period (102.9 days).  The ({\it V$-$R}$_C$) data were first seasonally normalized to remove a slight reddening trend in the more recent data.
\label{20056}}
\epsscale{1.0}
\end{figure}

The photometric results of this study confirm the classification of IRAS 20056$+$1834 as an RV Tau variable.  It shows the characteristics of alternating deep and shallow minima of variable depths and a secondary period (51.5 days) of half of the primary period (102.9 days), the latter of which is defined as the time between the deeper minima.  This is seen in the top panel of Figure~\ref{20056}, where the VUO-old and VUO-new {\it V} data are plotted in phase using the results of the combined {\it V} solution listed in Table~\ref{periods}, with the long period variation removed.  
The spectral type of G0~Ie is appropriate for an RV Tau variable.  The presence of a long secondary period would result in the sub-classification as RVb type  \citep{per07}.

Also included in Figure~\ref{20056} is the ({\it V$-$R}$_C$) color curve, plotted using the same ephemeris as used for the {\it V} light curve.  The color curve shows the same basic pattern of variability as the light curve, but with a shift (0.10) to an earlier phase.  This results in the star being bluest (or least red) on the rising branch of minimum light, and a little bit bluer on the rise from the shallower minimum, in accord with the general findings of \citet{pol96} in their photometric study of a sample of RV Tau variables.
The amplitude of {\it P}$_1$ is larger in the shorter {\it V} bandpass, while interestingly, the amplitudes are approximately the same for each of {\it P}$_2$ and {\it P}$_{\rm long}$ in the {\it V} and {\it R}$_C$ bandpasses.  

We also find evidence of a period change in the star.  A period between deeper minima of 102.9 days was found based on data from 2003 to 2018, while the period determined from the interval 2018$-$2022 yielded a period of 104.8 days.  Such changes are common in RV Tau variables \citep{per97}.

\subsection{How Common is a Long Period in the Light Curves of Post-AGB Objects?}   

In this study, we have documented the presence of a long period in the light curves of four objects classified as post-AGB stars.  
There is strong evidence from other studies that this is a signature of a binary companion, with the light modulation caused by the barycenter motion of the evolved star within a circumbinary disk.
As noted in the previous subsection, there exists a subclass of RV Tau variables which possess long-period (700$-$2500 d) modulation in their light curves, classified as RVb.
\citet{kiss07}, in a light curve study of 30 post-AGB stars based on ASAS-3 data, found approximately eight sources with long secondary periods that would be thus classified.
Most of these have appeared in studies of post-AGB disk sources \citep{deruy06,klus22}.
Radial velocity studies support the binarity of the RVb \citep{man17} variables.  
Binarity has also been suggested for red giants with long-period light curve modulations \citep{sosz21}.
Since the presence of a binary companion can be important in the evolution of the star and in the shaping of its ejected material, we would like to know how common it is in PPNe shell sources.
Is it the mechanism responsible for the bipolar morphology frequently seen in PPNe?
For one of these four, IRAS 08005$-$2356, the presence of a companion was recently confirmed based on radial velocity variations \citep{man21}.  

To begin to answer this question, we offer the results found in two samples.
In a series of papers studying the photometric variations in PPNe over 10$-$25 years \citep[and references therein]{hri22}, we analyzed the light curves of 29 PPNe in the Milky Way. 
In these studies, we have detected only three clear cases of long-period variations \citep{hri21}.  They are listed in the lower portion of Table~\ref{periods}.  
The results are not quite as robust as for the four in this study.  One (IRAS 14325$-$6428) has a large amplitude but the observations cover only one period and include a five year gap.  The other two have much smaller amplitudes, much smaller than the stellar variations seen in a single season.  All three of these appear to be valid results but all would benefit from continued light curve monitoring.

Secondly, we investigated the ASAS-3 light curves for the sample of 249 post-AGB candidates studied by \cite{oud22}.  Their sample was chosen for stars with ``likely'' or ``possible'' identification as post-AGB stars and which possessed Gaia DR3 parallaxes.  Since ASAS-3 only observes stars with decl. $\le$ $+$29$\arcdeg$, 
this eliminated many of these stars.
Also, there are cases with no ASAS-3 light curve and some cases in which the data are too few to make a judgement.  
This resulted in a reduced sample of 114 post-AGB candidates with ASAS-3 light curves.  We examined these ASAS-3 light curves visually and tentatively identified five as having ``definite'' and several more as having ``likely'' long-term cyclical variations.  
The five ``definite'' cases include two of our four objects, IRAS 07253$-$2001 and 08005$-$2356, and another object which has a measured radial velocity period of 1289 days \citep{oom18} and a SED indicating warm dust in a disk. 
After assessing these $-$ looking at the light curves, the SEDs (double peaked with a second peak in the near- or mid-infrared), and other known properties $-$ we judge that there may be six additional post-AGB candidates with long-term periods among this sample of 114 objects.
We note, however, that the precision of the ASAS-3 data is not good by present standards, with $\sigma$$_{\rm ave}$ = 0.04$-$0.07 mag.
For this reason, the long-period light variations of IRAS 17542$-$0603 and 20056$+$1834 were not recognized in the ASAS-3 light curve.  
Thus, there are likely some additional objects in this Oudmeijer et al. sample with long-term light variations that are below the level of ASAS-3 detection.  
However, many of them would be expected to be post-AGB disk objects and not PPNe shell objects. 

These samples, within their limitations, suggest that the number of likely PPNe objects with long-period variations due to a binary companion is a small fraction of the known PPNe candidates.  One of those we identified in the \citet{oud22} sample as ``definite'' and one as ``likely'' are indeed known binaries with disks \citep{klus22}.

\subsection{Binary Confirmation $-$ Radial Velocities}

An obvious way to confirm the binary nature of these objects would be through a radial velocity study.  
This approach has been very successful in identifying binaries in post-AGB objects with disks \citep{deruy06}.
Recently such a binary confirmation was made for IRAS 08005$-$2356.  \citet{man21} found a binary period of 7.3 yr based on its radial velocities and near-infrared light curves.
The radial velocity curve has a semi-amplitude of 8.2 km~s$^{-1}$ and an eccentricity of 0.36.
For IRAS 20056$+$1834, such an investigation is difficult. A spectroscopic study by \citet{kloch07} found variable radial velocities.
However, due to the evidence that the star is obscured by its circumstellar shell and seen primarily in scattered light, interpreting the radial velocities would not be straightforward.  Nevertheless, it would be worthwhile effort to monitor the radial velocities to investigate the presence of a periodicity to see if it is similar to that of the long period found in the light curve.
IRAS 07253$-$2001 and 17542$-$0603 are recommended for follow-up radial velocity monitoring, along with the three previously identified \citep{hri22}.  
These objects are faint, {\it V} $=$ 11.8$-$13.5, and red, so a medium-large telescope is required.    
The estimated semi-amplitudes are reasonable.  Assuming M$_{\rm PPNe}$ = 0.58 M$_\sun$, M$_{\rm 2}$ = 0.5$-$0.8 M$_\sun$ and a circular orbit, then the semi-amplitudes for IRAS 07253$-$0603 would be 8, 7, 4 km~s$^{-1}$ for binary inclinations of 90$\arcdeg$, 60$\arcdeg$ and 30$\arcdeg$, respectively. The values for IRAS 17542$-$0603, with its longer period, are 84$\%$ of these values.  These are similar to the value of 8.2 km~s$^{-1}$ found for IRAS 08005$-$2356 \citep{man21}.

\section{RESULTS AND SUMMARY}

In this study, we presented new light curves of four post-AGB stars, observed over an interval of 15$-$25 years.  We found the following:

1. All four displayed long-term light variations, with periods from 3.95 to 8.2 yrs.  These variations are suggested to be due to the periodic occultation of the evolved star by a circumstellar disk as the star moves in its barycentric orbit.

2. Three of these (IRAS 07253$-$2001, 08005$-$2356, and 17542$-$0603) appear to be PPNe.  A pulsation period of 73.6 days was found in IRAS 17542$-$0603.  
The other two varied in light, but no periodicity was found.

3. The fourth, IRAS 20056$+$1834, is found to be an RV Tau variable of the RVb subclass, with a formal period (from deeper minimum to deeper minimum) of 102.9 days and a secondary period of half that value (51.5 days).  It also showed evidence of a recent period change at the 2$\%$ level.

4. There is the suggestion in the light curve of IRAS 17542$-$0603, of a mass ejection event around or before 1994 that temporarily obscured some light from the star but at the same time selectively scattered blue light into our line of site.

One of the PPNe, IRAS 08005$-$2356 has recently been shown to be a binary based on a radial velocity study \citep{man21}.  
Radial velocity monitoring of PPNe with long-period light variations is recommended to test the binary model.
If binarity is established, then the physical properties can be used to constrain models for the shaping of the nebulae in these PPNe. 
If they are not found to be binaries, then we are back where we started, with a highly suggestive model for the shaping of the nebulae of PPNe, but with a paucity of direct evidence to support the binary nature of the objects.

\acknowledgments

We acknowledge with thanks the many Valparaiso University undergraduate students who participated in the observing of these stars at the Valparaiso University Observatory.
The ongoing technical support of Paul Nord is gratefully acknowledged.
Undergraduates Sean Egan, Peyton Grimm, and Matthew Bremer assisted in the analyses and the ASAS-3 light curve retrieval.  
We appreciate the comments of the referee, which especially helped to sharpen the motivation for the study and referred us to some additional relevant papers.
BJH acknowledges ongoing support from the National Science Foundation (most recently AST-1413660) and the Indiana Space Grant Consortium.
This research has made use of the SIMBAD and VizieR databases, operated at CDS, Strasbourg,
France, and NASA's Astrophysical Data System. 

\section{Appendix: Combining the Multi-telescope Observations}
\label{append}

As mentioned earlier, the SARA data reveal systematic offsets between the different camera-filter systems.  This was found in earlier studies and is seen again here. 
We attribute these primarily to the neglect of second-order terms in the standardization of these red stars.
We determined the individual offsets for each star with each camera empirically.  
This was done beginning with the average camera offset values from two previous studies and then by inter-comparison of the data for a particular star observed with the different camera systems.  For IRAS 17542$-$0603, for which we have observations from both VUO-new and the SARA cameras in the same seasons, we could determine the offsets directly for some of the cameras.  
A fuller discussion of these photometric offsets is given by \citet{hri20a, hri20b}. 
The offset values used are listed in Table~\ref{offsets}, and they are in reference to the VUO-new system.  
It is assumed that there is no offset between the VUO-old and VUO-new systems, but we are not able to check this directly as there were no simultaneous observations.  
No obvious systematic changes in brightness of the stars is obvious visually between the two VUO systems.
The offset values are almost all small, within a range of $\pm$0.02 mag.  However, for two of the cameras the offset values for IRAS 08005$-$2356 are large, $+$0.05 to $+$0.075 mag.  We don't think that these uncertainties impact the results of these studies.

Our photometric {\it V} and {\it R}$_C$ data for the three program stars that we observed, with the offsets included, are listed in Table~\ref{std_mags}, which is available in its entirety in machine-readable form.  Included are the heliocentric Julian date (HJD) of the observation, the standardized differential magnitude (program star $-$ comparison star 1), and a code to identify the particular telescope-detector-filter set used.
These codes are identified in Table~\ref{offsets}.

\facility{SARA)}
\facility{ASAS-3)}
\facility{ASAS-SN)}

\clearpage

\tablenum{1}
\begin{deluxetable}{crrrrrrrccl}
\rotate
\tablecaption{List of Post-AGB Objects Observed\label{object_list}}
\tabletypesize{\footnotesize} \tablewidth{0pt} \tablehead{
\colhead{IRAS ID}&\colhead{2MASS ID}&\colhead{Gaia ID}&\colhead{R.A.\tablenotemark{a}}&\colhead{Decl.\tablenotemark{a}}
&\colhead{{\it l}}&\colhead{{\it b}}
&\colhead{{\it V}\tablenotemark{b}}&\colhead{{\it B$-$V}\tablenotemark{b}}&\colhead{Sp.T.\tablenotemark{c}}&\colhead{Other ID}\\
&&&(2000.0)&(2000.0)&($\arcdeg$)&($\arcdeg$)&\colhead{(mag)} &\colhead{(mag)} & } 
\startdata
07253$-$2001 & 07273298$-$2007195 & 5620444471847839232 & 07:27:33.0 & $-$20:07:20 & 234.9 & $-$01.5 & 12.9 & 0.7 & F2I\tablenotemark{c}, F5I(e)\tablenotemark{d} & \nodata \\  
08005$-$2356 & 08024071$-$2404427 & 5698817012142459136 & 08:02:40.7 & $-$24:04:43 & 242.4 & $+$03.6 & 11.2 & 1.3 & F5~Iae\tablenotemark{c} & V510 Pup \\ 
17542$-$0603 & 17565602$-$0604096 & 4172337943816530432 & 17:56:56.0 & $-$06:04:10 & 021.2 & $+$09.2 & 12.7 & 1.4 & em\tablenotemark{c}, Ge\tablenotemark{e}, F6I\tablenotemark{f} & SS~336 \\ 
20056$+$1834 & 20075461$+$1842544 &1822346278591918336 & 20:07:54.6 & $+$18:42:54 & 058.4 & $-$07.5 & 12.4 & 1.1 & GOIe\tablenotemark{g} & QY Sge \\ 
\enddata
\tablenotetext{a}{Coordinates from the 2MASS Catalog.}
\tablenotetext{b}{These values are all variable as discussed in this paper.  They are based on our measurements, except the {\it B$-$V} measurements of IRAS 08005$-$2356 and IRAS 20056$+$1834 \citep{hen12} .}
\tablenotetext{c}{The spectral types from \citet{suarez06}.}
\tablenotetext{d}{The spectral type from \cite{red96}.}
\tablenotetext{d}{The spectral type from \citet{hu93}.}
\tablenotetext{e}{Spectral type by \citet{kel05}. }
\tablenotetext{f}{Spectral type by \citet{vie03}. }
\tablenotetext{g}{The spectral type from \citet{men88}.}  
\end{deluxetable}

\clearpage 

\tablenum{2}
\begin{deluxetable}{lccc}
\tablecaption{Observational Data Sets Used in this Study\label{data_sets}}
\tabletypesize{\footnotesize}
\tablewidth{0pt} \tablehead{\colhead{IRAS ID} &\colhead{Filter} 
&\colhead{Data Set} &\colhead{Dates}  } 
\startdata
07253$-$2001   & V & ASAS-3 & Nov 2000 $-$ Feb 2009   \\
 & V & ASAS-SN& Dec 2014 $-$ Nov 2018   \\
 & g & ASAS-SN & Apr 2018 $-$ Feb 2023   \\
08005$-$2356 & V & ASAS-3 & Nov 2000 $-$ Feb 2009   \\
 & V,R & SARA & Feb 2009 $-$ Mar 2018   \\
 & V & ASAS-SN& Feb 2016 $-$ Sep 2018   \\
 & g & ASAS-SN & Aug 2018 $-$ Feb 2023  \\
17542$-$0603 & V & VUO-old & Jul 1994 $-$ Jul 2007   \\
 & R & VUO-old & Jun 1995 $-$ Jul 2007   \\
 & V,R & VUO-new & Jul 2008 $-$ Jul 2018   \\
 & V,R & SARA &  Jun 2010 $-$ Apr 2018   \\
 & V & ASAS-SN & Jan 2015 $-$ Sep 2018   \\
 & g & ASAS-SN & Apr 2018 $-$ Mar 2023   \\
20056$+$1834 & V,R & VUO-old & May 2003 $-$ Sep 2007  \\     
 & V,R & VUO-new & Jul 2008 $-$ Jul 2018  \\    
 & V & ASAS-SN & Mar 2015 $-$ Aug 2018   \\
 & g & ASAS-SN & Apr 2018 $-$ Feb 2023   \\  
\enddata
\end{deluxetable}


\tablenum{3}
\begin{deluxetable}{lrrrrl}
\tablecaption{Observed Standard Magnitudes and Colors of the Program Stars
\label{std_ppn}}
\tabletypesize{\footnotesize} 
\tablewidth{0pt} \tablehead{ \colhead{IRAS ID} &\colhead{{\it V}} &\colhead{{\it B$-$V}} &\colhead{{\it V$-$R$_C$}}
 &\colhead{{\it R$_C$$-$I$_C$}} 
&\colhead{Date}  \\
 &\colhead{(mag)} &\colhead{(mag)} &\colhead{(mag)}  &\colhead{(mag)} &\colhead{} } 
\startdata
07253$-$2001 & 12.88  & 0.74 &  0.47\tablenotemark{a} & 0.49\tablenotemark{a} & 1993 Apr 8\tablenotemark{b} \\   
08005$-$2356  & 11.24   & \nodata  & 0.93\tablenotemark{a} & \nodata & 2013 Dec 27 \\           
17542$-$0603  & 13.37   & 1.44 & 0.88\tablenotemark{a} & 0.84\tablenotemark{a} & 1994 Jun 24\tablenotemark{c} \\
                         & 12.72  & \nodata & 0.93\tablenotemark{a}  & \nodata & 2014 Jun 06  \\
20056$+$1834 & 12.39   & \nodata &  0.63 & \nodata\tablenotemark{d} & 2003 Jul 24  \\    
\enddata
\tablecomments{Uncertainties in the brightness and color are $\pm$0.01$-$0.02 mag except for IRAS 17542$-$0603 on 1994 Jun 24, for which the values are $\pm$0.02$-$0.03 mag.}
\tablenotetext{a}{Note that the {\it R}$_C$ measurement contains a contribution from H$\alpha$ emission. }
\tablenotetext{b}{Observed earlier at Cerro Tololo Inter-American Observatory.}
\tablenotetext{c}{Observed at Kitt Peak National Observatory.}
\tablenotetext{d}{{\it V$-$I}$_R$ $=$ 1.20 mag.}
\end{deluxetable}


\tablenum{4}
\begin{deluxetable}{lllrrr}
\tablecaption{Comparison Star Identifications and Standard Magnitudes \label{std_comp}}
\tabletypesize{\footnotesize}
\tablewidth{0pt} \tablehead{\colhead{IRAS Field}
&\colhead{Object} &\colhead{2MASS ID} &\colhead{{\it V}} &\colhead{{\it V$-$R$_C$}}
&\colhead{{\it V$-$I$_C$}} \\
&\colhead{} &\colhead{} &\colhead{(mag)} &\colhead{(mag)} 
&\colhead{(mag)} } \startdata
08005$-$2356 & C1 & 08024410$-$2404362 & 12.47 & 0.70 & \nodata  \\
              & C2 & 08022747$-$2405103 & 12.57 & 0.60 & \nodata  \\
	      & C3 & 08023750$-$2406465 & 10.78 & 0.11 & \nodata  \\              
17542$-$0603 & C1 & 17565448$-$0604341 & 13.40 & 0.65 & \nodata  \\
              & C2 & 17564640$-$0608389 & 13.23 & 1.14 & \nodata  \\
              & C3 & 17562649$-$0605175 & 12.96 & 1.06 & \nodata   \\
20056$+$1834 & C1 & 20075619$+$1842482 & 11.76 & 0.67 & 1.28  \\
              & C2 & 20075820$+$1842308 & 12.36 & 0.98 & 1.98  \\
              & C3 & 20075287$+$1841072 & \nodata & \nodata & \nodata \\              
\enddata
\tablecomments{Uncertainties in the brightness and color are $\pm$0.01$-$0.02 mag.}
\end{deluxetable}

\clearpage

\tablenum{5}
\begin{deluxetable}{lcrrrrrrrrrrrrr}
\tablecolumns{17} \tabletypesize{\scriptsize}
\tablecaption{Results of the Periodogram Study of the Light Curves\tablenotemark{a}\label{periods}}
\rotate
\tabletypesize{\footnotesize} 
\tablewidth{0pt} \tablehead{ 
\colhead{IRAS ID} &\colhead{Filter} & \colhead{Years} & \colhead{Data\tablenotemark{b}} & \colhead{No.} & \colhead{{\it P}$_{\rm Long}$} & \colhead{{\it A}$_{\rm Long}$} & \colhead{{\it $\phi$}$_{\rm Long}$\tablenotemark{c}} & \colhead{{\it P}$_1$}&\colhead{{\it A}$_1$} &\colhead{$\phi$$_1$\tablenotemark{c}} & \colhead{{\it P}$_2$} & \colhead{{\it A}$_2$} & \colhead{$\phi$$_2$\tablenotemark{c}} & \colhead{$\sigma$\tablenotemark{d}} \\
 & & & \colhead{Sets} &\colhead{Obs.} & \colhead{(yr)}&\colhead{(mag)} & &\colhead{(day)}&\colhead{(mag)}
 & &\colhead{(day)} &\colhead{(mag)}&  &\colhead{(mag)} }
\startdata
\multicolumn{14}{c}{...................................................... Long-Period Results ................................... Pulsation Results} \\
\tableline 
07253$-$2001 & {\it V} & 2000-2018 & 2,5    & 738    & 5.2 & 0.13 & 0.20 & 73.6 & 0.069 & 0.21  & \nodata & \nodata & \nodata & 0.083 \\
08005$-$2356 & {\it V} & 2010-2018 & 2,4,5 & 1076 & 6.9 & 0.093 & 0.89 & \nodata & \nodata & \nodata  & \nodata & \nodata & \nodata & 0.045 \\
17542$-$0603 & {\it V} & 2003-2018 & 1,3,4,5 & 875 & 8.2 & 0.058 & 0.57 & \nodata & \nodata & \nodata  & \nodata & \nodata & \nodata & 0.032 \\
20056$+$1834 & {\it V} & 2003-2018 & 1,3,5 & 393 & 3.85 & 0.089 & 0.43 &102.9 & 0.097 & 0.96  & 51.5 & 0.070 & 0.57 & 0.049 \\
20056$+$1834 & {\it g} & 2018-2023 & 5 & 509 & \nodata & \nodata & \nodata &104.8 & 0.143 & 0.40  & 52.4 & 0.055 & 0.50 & 0.054 \\
\tableline 
\multicolumn{14}{c}{Long-Period Results $-$ Previous Study\tablenotemark{e}} \\
\tableline 
08143$-$4406 & {\it V} & 2000-2018 & 1,2,3 & 858 & 5.0 & 0.021 & 0.79 & \nodata & \nodata & \nodata & \nodata & \nodata & \nodata & 0.051 \\
14325$-$6428& {\it V} & 2000-2018 & 1,2,3 & 790 & 18.8 & 0.096 & 0.41 & \nodata & \nodata & \nodata & \nodata & \nodata & \nodata & 0.041 \\
15482$-$5741 & {\it V} & 2000-2018 & 1,2,3 & 583 & 9.6 & 0.037 & 0.93 & \nodata & \nodata & \nodata & \nodata & \nodata & \nodata & 0.053 \\
\enddata
\tablenotetext{a}{The uncertainties in the parameters for the four program stars of this study are as follows: period ({\it P}) $-$ $\pm$0.01$-$0.1 days for pulsation and $\pm$0.1 yr for the long-periods except $\pm$0.05 yr for IRAS 20056$+$1834; amplitude ({\it A}) $-$  $\pm$0.002$-$0.004 mag; phase ($\phi$)  $-$ $\pm$0.01$-$0.03. }
\tablenotetext{b}{1 = VUO-old, 2 = ASAS-3, 3 = VUO-new, 4 = SARA, 5 = ASAS-SN.}
\tablenotetext{c}{The phases are determined based on the epoch of 2,455,600.0000, and they each represent the phase derived from a sine-curve fit to the data, not the phase of minimum light.}
\tablenotetext{d}{Standard deviation of the observations from the sine-curve fit.}
\tablenotetext{e}{Results from the study by \citet{hri21}.}
\end{deluxetable}

\clearpage

\tablenum{A1}
\begin{deluxetable}{lcllllllllll}
\tablecaption{SARA Telescope-Detector-Filter Offsets for Each Star\tablenotemark{a}\label{offsets}}
\tabletypesize{\footnotesize} \tablewidth{0pt} \tablehead{
\colhead{Telescope-}&\colhead{Code\tablenotemark{b}}&&\multicolumn{2}{c}{IRAS 08005}&&\multicolumn{2}{c}{IRAS 17542} &&\multicolumn{2}{c}{IRAS 20056} \\
\cline{4-5} \cline{7-8} \cline{10-11}
\colhead{Detector}&\colhead{}&&\colhead{{\it V}}&\colhead{{\it R}$_C$}&&\colhead{{\it V}}&\colhead{{\it R}$_C$} &&\colhead{{\it V}}&\colhead{{\it R}$_C$} \\
\colhead{}&\colhead{}&&\colhead{(mag)}&\colhead{(mag)}&&\colhead{(mag)}&\colhead{(mag)} &&\colhead{(mag)}&\colhead{(mag)}}
\startdata
 VUO-new & A   &&  \nodata  &  \nodata  && $+$0.00 & $+$0.00 &&  $+$0.00 & $+$0.00    \\
 VUO-old  & B   &&  \nodata  &  \nodata   && $+$0.00\tablenotemark{c} & $+$0.00\tablenotemark{c} &&  $+$0.00\tablenotemark{c} & $+$0.00\tablenotemark{c}    \\
SARA-KP U42   & C  && $-$0.010 & $+$0.015 && $-$0.01 & $-$0.01 &&  \nodata  &  \nodata      \\
SARA-KP ARC  & D  && $+$0.00 & $+$0.01 && $+$0.00: & $+$0.02: &&  \nodata  &  \nodata   \\
SARA-CT E6     & E && $+$0.07:  & $+$0.075: && $+$0.035 & $+$0.015 &&  \nodata  &  \nodata   \\
SARA-CT ARC & F && $+$0.00 & $+$0.005 && $+$0.02 & $+$0.01 &&  \nodata  &  \nodata   \\
SARA-CT FLI   & G && $+$0.05 & $+$0.05: && $-$0.02: & $+$0.01: &&  \nodata  &  \nodata   \\
\enddata
\tablecomments{Uncertainties are $\pm$0.01 to $\pm$0.02 mag; those with colons (:) are the more uncertain.  No observations were made for IRAS 08005$-$2356 with the VUO-new or VUO-old system.}
\tablenotetext{a}{Offsets as compared to the VUO-new values, in the sense that the offset value = mag(VUO-new) $-$ mag(system).  
The offset values were then added to the SARA magnitudes for the combined analyses. }
\tablenotetext{b}{Code is used to identify the source of the photometry and the associated offset to be used with the data in Table~\ref{std_mags} to bring the observations to the VUO-new system. }
\tablenotetext{c}{Offset assumed to be $+$0.00 for VUO-old. }
\end{deluxetable}


\tablenum{A2}
\begin{deluxetable}{llclcc}
\tablecaption{Differential Standard Magnitudes\label{std_mags}}
\tablewidth{0pt} \tablehead{\colhead{IRAS ID}
&\colhead{HJD({\it V})} &\colhead{$\Delta${\it V}} &\colhead{HJD({\it R}$_C$)} &\colhead{$\Delta${\it R}$_C$} 
&\colhead{Code\tablenotemark{a}} \\
\colhead{} &\colhead{} &\colhead{(mag)} &\colhead{} 
&\colhead{(mag)} & \colhead{}} 
\startdata

IRAS08005-2356  &  54890.7141  &  -1.328  & 54890.7171  &  -1.558  &  C \\
IRAS08005-2356  &  54901.6354  &  -1.368  & 54901.6360  &  -1.560  &  C \\
IRAS08005-2356  &  54908.6796  &  -1.346  & 54908.6791  &  -1.546  &  C \\
IRAS08005-2356  &  54921.6729  &  -1.335  & 54921.6722  &  -1.533  &  C \\
IRAS08005-2356  &  54924.6213  &  -1.319  & 54924.6166  &  -1.536  &  C \\
IRAS08005-2356  &  54928.7349  &  -1.330  & 54928.7415  &  -1.538  &  C \\
IRAS08005-2356  &  54944.6170  &  -1.341  & 54944.6140  &  -1.503  &  C \\
IRAS08005-2356  &  54945.6719  &  -1.342  & 54945.6709  &  -1.673  &  C \\
IRAS08005-2356  &  55146.9934  &  -1.313  &             &          &  C \\
IRAS08005-2356  &  55160.9463  &  -1.349  & 55160.9425  &  -1.554  &  C \\
\enddata
\tablecomments{Typical uncertainty in the standardized differential magnitudes is $\pm$0.015 mag.}
\tablenotetext{a}{This identifies the telescope-detector system used, as listed in Table~\ref{offsets}, and the associated added offset, if any, to bring that observation to the VUO-new system. \\ (This table is available in its entirety in machine-readable form.)}
\end{deluxetable}

\clearpage

\clearpage

\clearpage

\end{document}